\documentclass[
reprint,
groupedaddress,
amsmath,
amssymb,
aps,
prb,
longbibliography,
noeprint,
raggedbottom
]{revtex4-2}

\usepackage{txfonts}
\usepackage{epsfig}
\usepackage{dcolumn}
\usepackage{bm}
\usepackage{color}
\usepackage{graphicx}
\usepackage{bbold}

\begin{document}

\title{Classical Analog Circuit Emulation of Quantum Grover Search Algorithm}

\author{Samuel Feldman} 
\affiliation {Department of Physics and Astronomy, University of Utah, Salt Lake City 84093, USA}
\author{Hassam Ghazali} 
\affiliation {Department of Physics and Astronomy, University of Utah, Salt Lake City 84093, USA}
\author{Andrey Rogachev}
\affiliation {Department of Physics and Astronomy, University of Utah, Salt Lake City 84093, USA}

\date{\today}

\begin{abstract}
We construct a completely analog framework that emulates universal quantum gates and quantum algorithms. It is based on electronic circuits made of operational amplifiers, resistors and capacitors. In these circuits, input and output lines represent the computational basis states (CBSs) and thus $2^n$ lines are required to represent $n$ qubits. An operation of the circuits is based on classical evolution and interference of complex amplitudes associated with each CBS. The framework can emulate entangled states and is free from decoherence, measurements are classical and do not collapse states. Similar to physical quantum computers, emulated quantum algorithms can be constructed as a sequence of the gates belonging to a universal set (phase shift, Hadamard, controlled-NOT), as a unitary matrix and as combinations of the two. Circuits representing the universal gates have been made and tested.  We also have made a matrix-based emulator of a 3-qubit Grover’s search algorithm.  We tested it by searching for one and two particular states with one and two iterations and found that its outputs accurately match predicted values.  We anticipate that the emulators can work as sub-components of physical quantum computers. On their own, the emulators can be used for operations that require a few qubits or operations that can be split into independent (and perhaps weakly entangled) blocks of qubits.
\end{abstract}

\maketitle

\noindent\textbf{I. Introduction}

Quantum computers have the potential to perform computations that are intractable for classical computers. This advantage is achieved via \textit{quantum parallelism}, by which a system consisting of $n$ individual qubits can operate on the $2^n$ separate combined states of those qubits simultaneously. The problems quantum algorithms are predicted to speed up include factoring large numbers \cite{1}, drug discovery \cite{2}, material design \cite{3,4,5}, and search and optimization problems \cite{6,7}. Qubits can be implemented on any quantum 2 level system, with the most promising current candidates being superconducting circuits \cite{8}, neutral atoms \cite{9}, and trapped ions \cite{10}. Any possible quantum algorithm can be represented as a combination of a small set of 1 and 2-qubit gates, known as a universal gate set.

There are several barriers to physically realizing quantum computing, most notably the limited coherence time of qubits and the difficulty of entangling large numbers of qubits. The size and accuracy of physical quantum computers are increasing every year, and we have recently seen claims of quantum computers that outperform classical computers \cite{6,11}. However, this success has only been found on very specific problems, and the most intriguing applications of quantum computers remain beyond the reach of available systems.

Remarkably, it was recently realized that several quantum effects that are very difficult to observe in nature can be realized in classical analog circuits \cite{12}. These specialized circuits have successfully demonstrated a variety of topological effects including phase transitions \cite{13,14}, bulk-boundary correspondence \cite{15}, Majorana modes \cite{16},  and Weyl particles \cite{17}. This mapping is not limited to quantum mechanics, and similar circuits are used to study a massive range of complex systems including the human brain \cite{18,19,20,21}, evolution \cite{22}, and macroeconomics \cite{23}.

Concurrent with this exciting development, in several recent theoretical papers, Motohiko Ezawa has shown how classical circuits can be used to implement the universal quantum gate set \cite{24,25,26}. In Ezawa’s proposal, signals are carried by semi-infinite transmission lines and the gate operations are performed by inductive and capacitive linkages between the lines, acting as scattering elements. By carefully tuning the frequency, reflection can be minimized and the transmitted signal will reflect the output of the gate.

In this paper, we present an analog circuit framework emulating the action of the universal quantum gates and show how it can be used to implement Grover’s Search Algorithm (GSA) \cite{6}. While we follow Ezawa’s approach conceptually, our hardware implementation is different and is based on circuits composed of operational amplifiers (op-amp), resistors, and capacitors.  Our circuits are lumped, so their operation is not affected by transmission and reflection. The addition of active elements (op-amps) allows for introducing gain, so signals can be amplified or restored, and dissipation is no longer an issue. Our approach is close to an analog emulator proposed and simulated on a CMOS-based architecture in Ref. \cite{27}.  Also, there are several previous works on hardware emulation of quantum computers using digital field programmable gate arrays \cite{28,29,30}. 

\begin{figure*}[tb]
\includegraphics[width=1.0\textwidth]{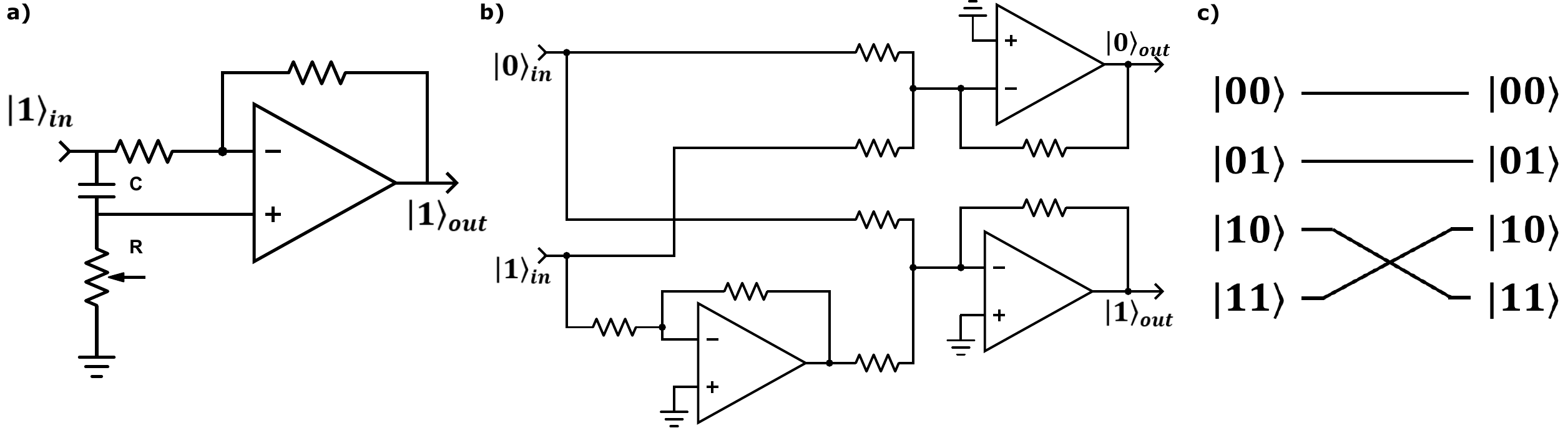}
\caption{\textbf{Circuit designs for the universal gate set.} 
\textbf{(a)} Phase-shift gate, implemented with a single op-amp. \textbf{(b)} Hadamard gate, implemented with three op-amps. \textbf{(c)} Controlled-Not gate.}
\label{f1}
\end{figure*}
 
\noindent\textbf{II. Quantum Gates}

Let us now describe our approach in detail. In a physical quantum computer, each qubit is a separate quantum system with 2 basis states, $|0\rangle$ and $|1\rangle$. The state of the qubit is described by its wavefunction, which is constructed of these two states as $\left|\Psi\right\rangle=\alpha\left|0\right\rangle+\beta|1\rangle$, where $\alpha$ and $\beta$ are the complex probability amplitudes of the qubit being in their corresponding state. It is common to combine these numbers into a 2-dimensional state vector that fully describes the state of the qubit. 

The fundamental barrier to classically emulating a quantum computation is the lack of entanglement scaling. Two entangled qubits cannot be described by their individual state vectors. Instead, we need to construct a vector with 4 elements, one for each possible combination of the individual qubit basis states 
\begin{equation}
\Psi=\alpha|00\rangle+\beta|01\rangle+\gamma|10\rangle+\delta|11\rangle
\end{equation}
These are the type of states used in classical computers to represent a quantum system.  In this paper, we will refer to the states like $|00\rangle,|01\rangle,|10\rangle,|11\rangle$ as computational basis states (CBSs) to make clear the distinction from individual qubit states. Emulating a system of $n$ entangled qubits requires storing both magnitude and phase for all $2^n$ possible CBSs. 

In our hardware design, each CBS is represented by an electrical line (coaxial cable or just a wire). The quantum amplitudes, such as $\alpha$ and $\beta$ for a single qubit, and $\alpha,\beta,\gamma,\delta$ for a 2-qubit system, are represented by AC signals characterized by their amplitudes and relative phase shifts.
The gates that can be implemented on a physical quantum computer are quite limited. Quantum mechanics requires that overall probability is conserved, so only unitary gates are allowed. In addition, the interaction of a quantum system with its environment promotes decoherence, so gates must be carefully implemented to reduce this chance. In comparison, circuit emulation is extremely flexible in how gates can be implemented. All quantum gates have a corresponding matrix, and we can use analog computing techniques to implement gates through matrix algebra. We can also use the matrix representations of individual gates that are applied consecutively to combine them into a single physical gate, reducing the number of components required. There is no need to implement all gates unitarily. As long as the relative amplitudes of each line are preserved, the final output can simply be renormalized.

The universality of circuit emulation is demonstrated by creating a universal gate set. We chose 3 gates: the single qubit arbitrary phase-shift and Hadamard gates, and the 2-qubit controlled-NOT gate. The phase-shift gate applies a shift to the $|1\rangle$ state of the qubit relative to the $\left|0\right\rangle$ state. A controllable, constant amplitude phase shifter can be constructed via an op-amp, as shown in Fig. 1(a). The phase shift can be controlled via the adjustable resistor and has a value of $\varphi=\pi-2\arctan{\left(2\pi fRC\right)}$, which varies between 0 and $\pi$. To realize a phase shift of greater than $\pi$, the variable resistor and capacitor should exchange their positions in the circuit. The $|0\rangle$ state signal is unchanged.

The emulator of the Hadamard gate composed of inverting and summing operational amplifiers, as shown in Fig. 1(b), implements the Hadamard matrix
\begin{equation}
H=\frac{1}{\sqrt{2}}\begin{bmatrix}1&1\\1&-1\end{bmatrix}
\end{equation}
This will turn the basis state $|0\rangle$ into the superposition state  $\left(\left|0\right\rangle+\left|1\right\rangle\right)/\sqrt2$. On the circuit diagram of Fig. 1(b), this means that if the input signal is applied only to $\left|0\right\rangle_{in}$, the outputs, marked as $\left|0\right\rangle_{out}$ and $\left|1\right\rangle_{out}$,  assume the same amplitude and phase. The basis state $\left|1\right\rangle$ evolves into a state $(\left|0\right\rangle-\left|1\right\rangle)/ \sqrt2$, meaning that the output $\left|1\right\rangle_{out}$ is $\pi$-shifted with respect to $\left|0\right\rangle_{out}$. The outputs of the Hadamard gate are the eigenvectors of the Pauli X matrix, so the Hadamard gate provides a change of basis between the Z and X bases.

The controlled-Not gate acts to invert the state of the target qubit when the control qubit is in the $\left|1\right\rangle$ state. The action on the target qubit is to swap amplitudes of its $|0\rangle$ and $|1\rangle$ states, which can be done by simply wiring the state to one another. In Fig. 1(c) the qubit that has the second position in the CBS notations is the control and the first qubit is the target. It is conventional to number qubits in a combined state from right to left to facilitate mapping to classical binary numbers. 

\begin{figure}[tbph]
\includegraphics[width=0.5\textwidth]{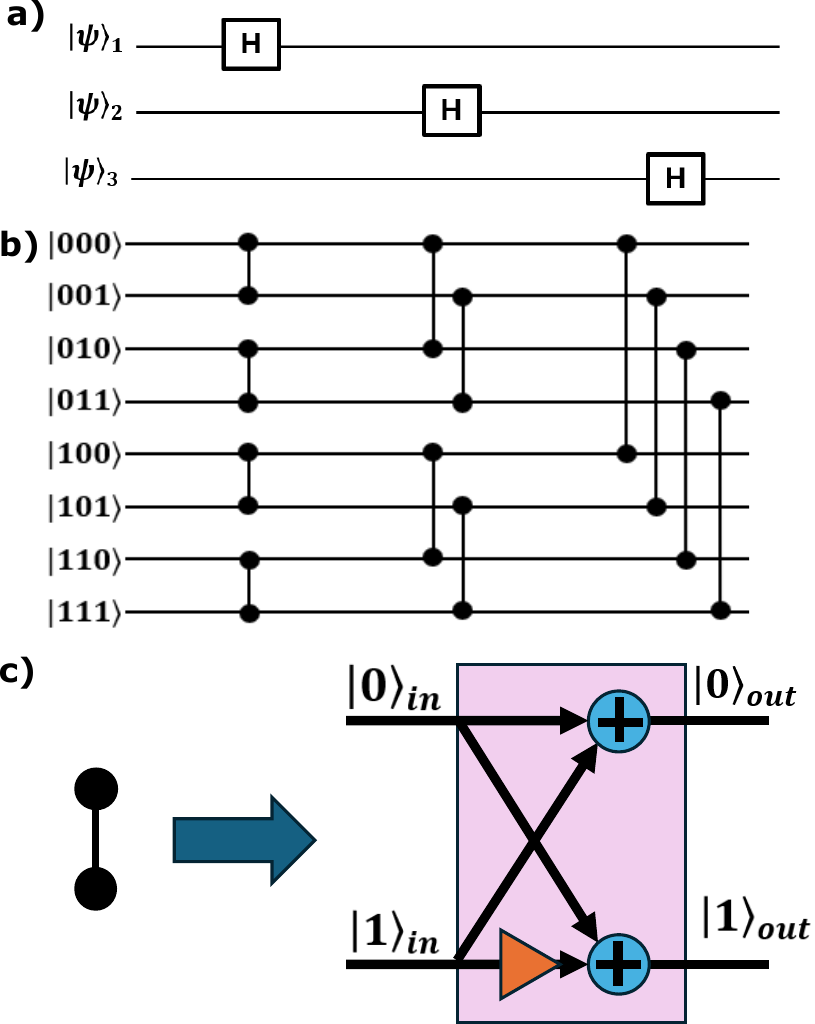}
\caption{\textbf{Application of Hadamard gate.} 
\textbf{(a)} Hadamard gate applied to each of the three qubits as in a physical quantum computer  \textbf{(b)} The same circuit but applied to a classical system  \textbf{(c)} Detail of the individual elements in \textbf{(b)}, the red triangle represents an inverter and blue circles are summers}
\label{f2}
\end{figure}

The circuits for a gate do not fully describe the implementation. Because our emulator uses CBSs as its basic elements, we cannot manipulate individual qubits directly. In some cases, a single qubit gate requires modifying all CBSs. As an example, Figs. 2(a) and (b) show equivalent quantum and classical circuits to change the basis of each of the three qubits via a Hadamard gate. To replicate a Hadamard gate on a single qubit, we create pairs of CBSs which have opposite states for the active qubit, but the same state for all other qubits. To replicate the action of three Hadamard gates, all the CBSs are paired up as shown in Fig. 2(c). The actual action of the gate on two paired CBSs (indicated in the circuit as two bold dots connected by a vertical line) is given by the Hadamard gate shown in Fig. 1(b).

The upside to the CBS scheme is that multiple qubit gates are often easy to implement. A controlled gate only affects the target qubit if the control qubit is in a certain state (usually $|1\rangle$). To control a gate by one or more qubits we only apply it to the CBS lines where the control qubit(s) is in the desired state. This is evident from Fig. 1(c). Similar simplification also occurs, for example, in a popular 3-qubit Toffoli (CCNOT) gate, which requires just a flip of two wires in the emulator \cite{25}, while the implementation on physical quantum computers requires several CNOTs.

We have built the quantum gate emulators presented in Fig. 1(a,b) and Fig. 2(b) using prototype circuit boards. All op-amps used are Analog Devices OP270. The operation of the circuits was tested in a continuous mode with an Agilent 33210A 10 MHz function generator providing a 1 kHz sine wave. Outputs were measured using an Agilent DSO1002A oscilloscope. The measured amplitudes and phase shifts of the outputs were found to match the expected values. In particular, we tested the case where the initial state of each qubit in Fig. 2(a) is $|0\rangle$. In the CBS representation of Fig. 2(b), this means that the state $|000\rangle$ has amplitude 1 and the seven other CBS states have amplitude 0.  We have found that after passing through the circuit, all CBS outputs assume, as expected, the same amplitude with no phase shift. 

\noindent\textbf{III. Grover's Search Algorithm}

Let us move now to the second part of the paper, constructing and testing a 3-qubit emulator of the Grover’s Search Algorithm (GSA). A very detailed description of this algorithm, both in the CBS representation and in the representation of physical qubits is given in Ref. \cite{31}. In the original formulation \cite{6}, GSA is proposed to use an $n$-qubit quantum computer to find a single special element in the array of $N=2^n$ classical elements in time $O\left(\sqrt N\right)$. Optimal classical search algorithms scale linearly. There is also an extension of GSA, known as amplitude amplification, that allows multiple elements to be searched for and marked \cite{7}. GSA’s performance begins to suffer if more than 1/4 of the possible states are special. 

Since our framework emulates quantum algorithms in the CBS representation, we discuss GSA in this representation first, with a concurrent description of the hardware implementation. The physical construction of our emulator consists of an aluminum chassis with 16 BNC connectors, one input and one output connector for each CBS. There are a series of solderless breadboards attached to the chassis. The circuit for each gate is soldered onto a perforated prototype board. Pins are attached to the ends of the boards, allowing them to be plugged into the breadboards and connected to each other. The completed circuit is shown in Fig. 3. 

GSA consists of three basic steps. First, all qubits in the system must be placed in an equal superposition. The overall quantum state of the system now has an equal probability amplitude in each possible combined state. This step is only performed once, while the next two steps may need to be applied several times. In a physical quantum computer, this is accomplished by applying a Hadamard gate to each qubit initially in $|0\rangle$ state. This action could be emulated by the circuit shown in Fig. 2, but we have skipped this step and produced its output directly by feeding each CBS line with the same AC signal passing through adjustable voltage dividers added to each line and set to a 50/50 ratio. 
\begin{figure}[b]
\includegraphics[width=0.5\textwidth]{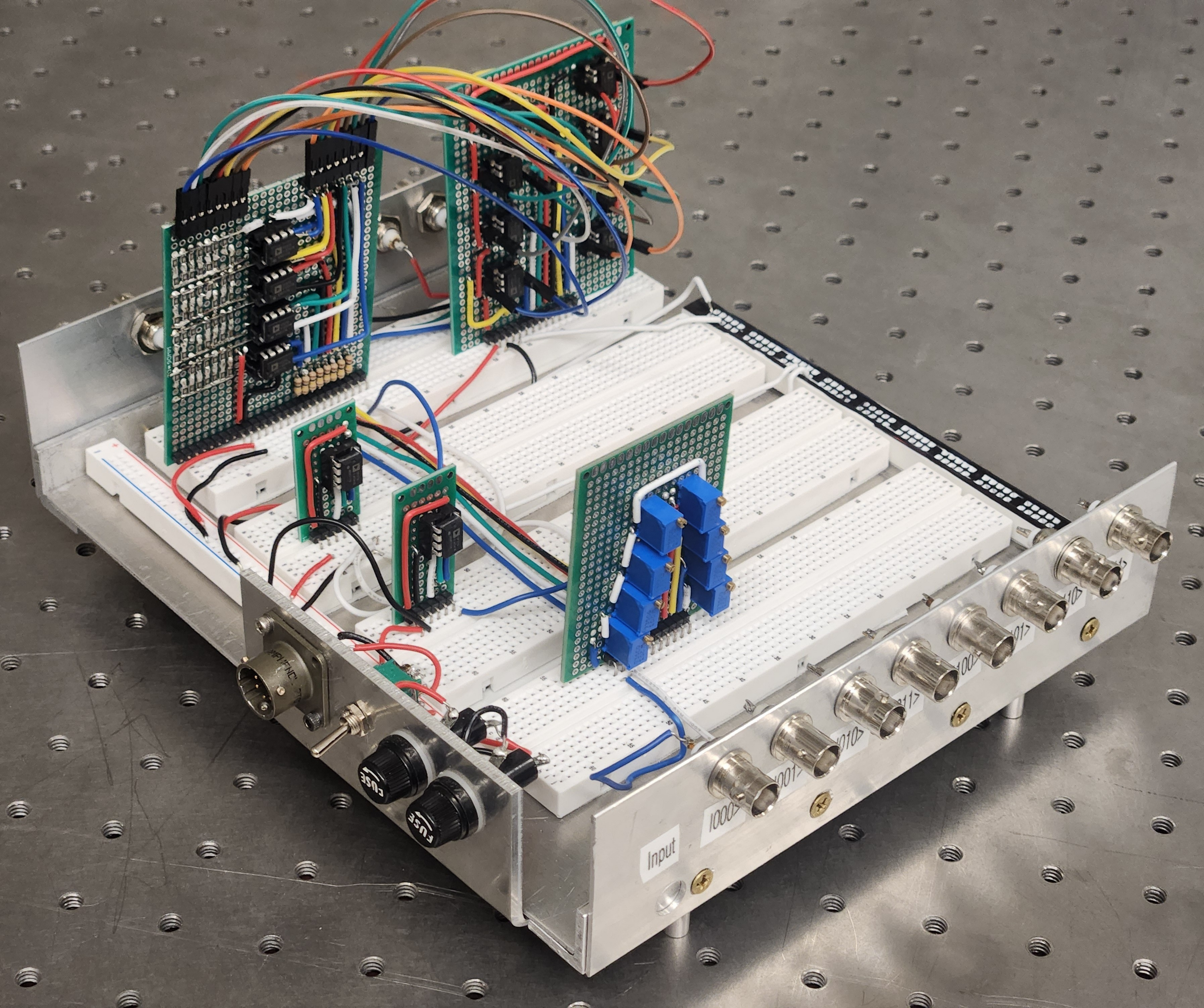}
\caption{\textbf{Classical circuit emulator for Grover's search for 2 special states.} 
Inputs ports for each CBS are at front right. The $|000\rangle$ input feeds the voltage divider gate which produces the equal superposition. The next two chips are the two inverters which implement the oracle operator. The large chips at the back implement the diffusion operator, where the left chip takes care of scaling and the right chip of summation. The outputs for each CBS are on the back plate of the emulator.}
\label{f4}
\end{figure}

The second step of GSA is to mark the desired state(s) by introducing a phase shift of $\pi$. In a real algorithm, this is supposed to be done by some external process (called an oracle) that communicates with an external system under search and marks a CBS unknown to GSA.  In our implementation we simply add inverters (as shown in Fig. 1 (a)) to one or two CBS lines and test if the marked state reveals itself in the following steps. 

The third step of GSA inverts the amplitude of every state around the mean amplitude. The matrix form of this gate on 3 qubits is

\begin{equation}
\frac{1}{4}
\begin{bmatrix}
  -3 & 1 & \cdots & 1 \\
  1 & -3 & \cdots & 1 \\
  \vdots & \vdots & \ddots & \vdots \\
  1 & 1 & \cdots & -3 
\end{bmatrix}
\end{equation}

\begin{figure}[b]
\includegraphics[width=0.5\textwidth]{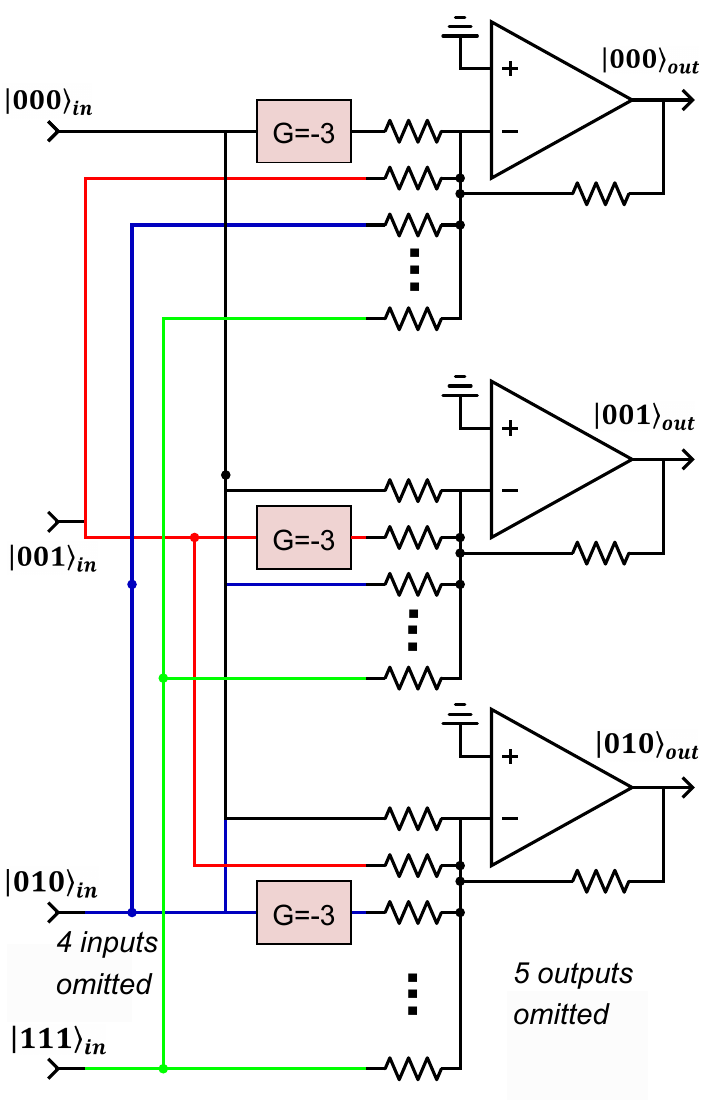}
\caption{\textbf{Truncated circuit diagram for the Grover Diffusion Gate.}}
\label{f3}
\end{figure}

Rather than breaking this operation into elementary gates, we implement it as a matrix with the equivalent circuit shown in Fig. 4. In this circuit, the signal in each line is reduced to 1/4 by voltage dividers and then the lines corresponding to the diagonal elements are put through an amplifier with gain of -3. We then use 8 op-amp summers, one representing each CBS, to add up the 8 contributions to each output state. This gate is also known as the Grover diffusion operator, because it implements a quantized analog of classical diffusion. The phase of the marked state(s) acts as an effective negative population, so the marked state(s) attract “particles” away from the more populous unmarked states. 
\begin{figure}[t]
\includegraphics[width=0.5\textwidth]{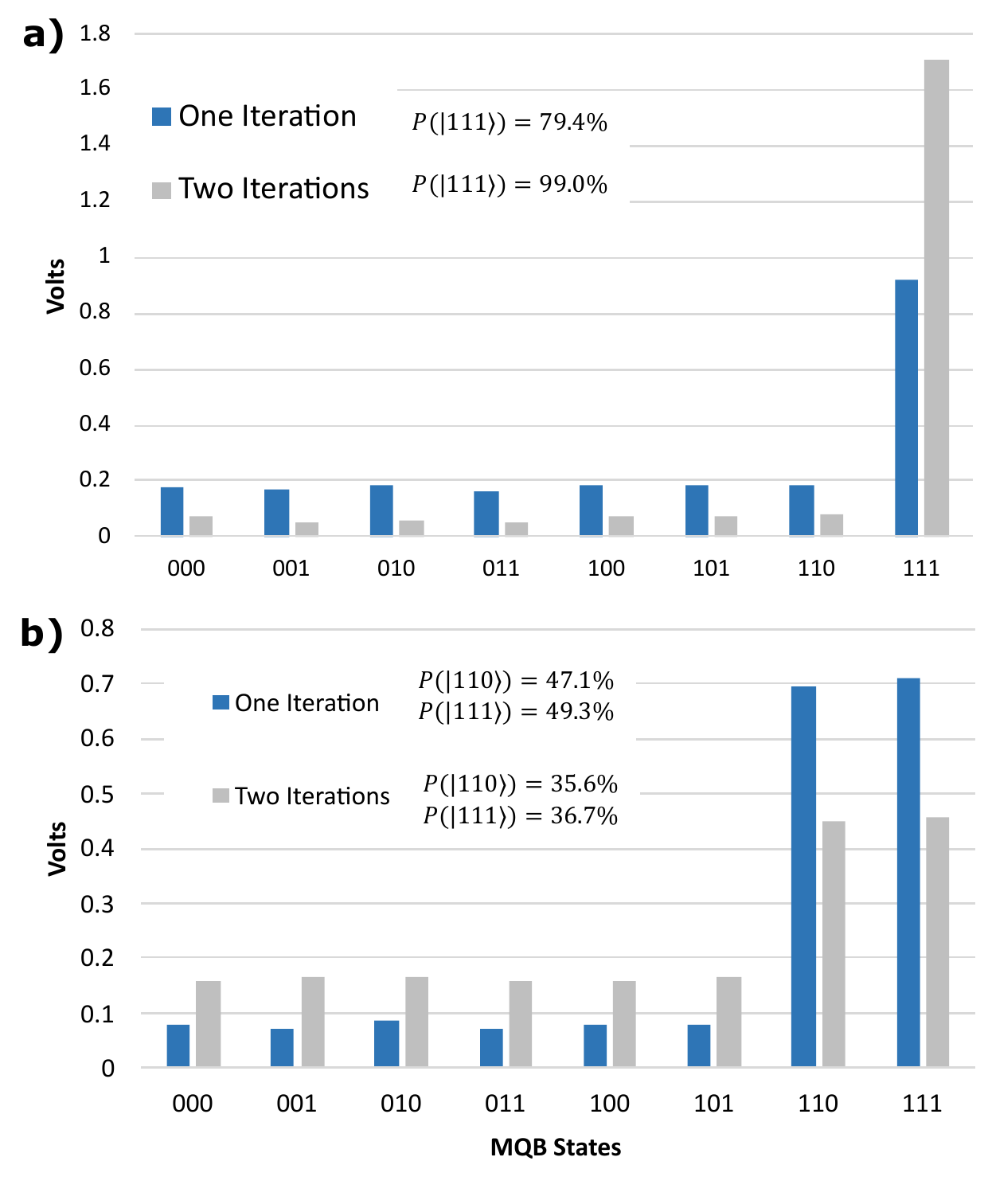}
\caption{\textbf{Measured output voltages for Grover’s Search Algorithm.} 
\textbf{(a)} Search for a single special state.  \textbf{(b)} Search for two special states}
\label{f5}
\end{figure}
The final step of any quantum algorithm is to perform a measurement. In the emulator, we do not have to worry about measurement induced collapse. We operated the circuit in a continuous mode and all the measurements have been done with the oscilloscope.  

After a single iteration of the Grover algorithm, the amplitude of the special states increases above average, and that of non-special states goes below average. In order to amplify this ratio, the second and third steps of the algorithm need to be iterated with inputs taken from the previous iteration. The theory states that for a search over N total CBS states with k special states, the optimal number of iterations is $\approx\left(\frac{\pi}{4}\right)\sqrt{\frac{N}{k}}$ \cite{7}. For 8 total states, a search for a single state is optimal at 2 iterations, and a search for 2 states is optimal at 1 iteration. We ran a second iteration of the algorithm by adjusting the voltage dividers so that the initial state matched the output of the first iteration. The ability to reuse a single physical gate multiple times in a circuit is another notable advantage of hardware emulation. 

Our measured results of the emulated GSA circuit are shown in Fig. 5. We compare results based on the Algorithm Success Probability (ASP), the sum of the probabilities of each special state. When searching for a single element, our ASP is 79.4\% after one iteration and 99.0\% after two. The outcome of the search for 2 special states has reached an optimal value after the first iteration with an ASP of 96.4\%. We thus found that our emulator fulfills all quantitative predictions of the Grover search algorithm. The tolerance of the components is apparently a source of residual errors. For completion, we also tested a second iteration of the search for 2 special states and found that the ASP dropped to 72.3\% as indeed expected when the iterations go beyond the optimal number. 

\noindent\textbf{IV. Summary and Discussion}

The presented experiments should be considered as proof of concept tests. It is technically straightforward to make these emulations more robust by automating data input and output signals and by using pulsed sine signals rather than continuous mode measurements. Moreover, unlike current physical quantum computers, this entire emulator is compatible with standard semicoductor technology and can be built on a single IC. In fact, this is the approach simulated in Ref. \cite{27}. In this work, each computational basis state is encoded with two lines (while we use just one). The authors of the paper estimated that their analog circuit produces a 400x speed-up compared to the corresponding classical digital simulation.  

Let us now briefly review the status of implementation of the GSA in physical quantum computers. Introducing a test oracle, which has been a straightforward step in our emulator, is a nontrivial task for a physical QC. It is proposed to be done by an ancilla qubit acting as the target for a controlled-X by all register qubits as well as a couple of additional X-gates in several qubits to encode the CBS  masked state on the $n$-qubit register state. For details please see Fig. 6.16 in Ref. \cite{31}.

Each additional qubit complicates the oracle design and increases the susceptibility to error. There are only a few studies that have managed to implement GSA with more than 3 qubits, and these are restricted to searching for a single state. Ref \cite{32} gives detail on designing 4 and 5 qubit oracles, each with a single ancilla qubit, and tests them on IBM quantum computers operating with superconducting qubits. The 4-qubit search was implemented with two Grover iterations and produced a maximum ASP of 19.\%. The 5-qubit search could only be run for a single iteration and had a max ASP of only 2.6\%. In a later paper \cite{33}, the 5-qubit search was implemented with 2 iterations on a Honeywell (now Quantinuum) H1 computer using trapped ion qubits. On this system the single-state ASP was found to be 49\%.

In summary, we have designed and built a system that emulates a universal quantum computer. This system is capable of accurately emulating any possible quantum algorithm, with the main drawback being the exponential increase in the number of lines needed to emulate an additional qubit. We have also successfully emulated a 3-qubit version of Grover’s Search algorithm demonstrating greater accuracy with less computational complexity than the current physical quantum computers.  

\smallskip
\noindent \textbf{Acknowledgements.}
The authors thanks Motohiko Ezawa for useful comments on his work. This research has been supported by NSF grant: DMR-2133014 

\smallskip
\noindent \textbf{Author Contributions.}
S.F. designed, built, and tested the circuit emulating the Grover algorithm. H.G. designed and tested the emulators of quantum gates. A.R. proposed the original concept of the analog framework and supervised the project. S.F. and A.R. wrote the manuscript. All authors contributed to discussions and reviewed the final version of the paper.

\smallskip
\noindent \textbf{Competing interests.}
The authors declare no competing interests

\smallskip
\noindent \textbf{Correspondance and requests for materials} should be addressed to Samuel Feldman.

\end{document}